\documentclass{lmcs} 


\usepackage{hyperref}
\usepackage{siunitx}
\graphicspath{ {./images/} }
\theoremstyle{plain} 

\begin{document}

\title[WSN's data synchronization using node MCU memory for Precision Agri. App's.]{Wireless sensor networks data synchronization using node MCU memory for precision agriculture applications}

\author[K.~Sattar]{Kashif Sattar\lmcsorcid{0000-0001-8620-9419}}[a]
\author[M.~Arslan]{Muhammad Arslan}[a]
\author[S.~Majeed]{Saqib Majeed}[a]
\author[S.~Iqbal]{Salim Iqbal}[b]
\address{University Institute of IT, PMAS-Arid Agriculture University Rawalpindi, Pakistan}	
\email{kashif@uaar.edu.pk, muhammadarslan5157@gmail.com, saqib@uaar.edu.pk}  

\address{Department of Computer Science, Allama Iqbal Open University, Islamabad, Pakistan}	
\email{saleem.iqbal@aiou.edu.pk}  





\begin{abstract}
  \noindent Wireless Sensor Networks have risen as a highly promising technology suitable for precision agriculture implementations, enabling efficient monitoring and control of agricultural processes. In precision agriculture, accurate and synchronized data collection is crucial for effective analysis and decision making. One of the important challenge is to achieve accurate data synchronization among sensor nodes in WSNs for precision agriculture applications. Using principles of information theory, we can define conditions and parameters that influence the efficient transmission and processing of information. Existing technologies have limitations in maintaining consistent time references, handling node failures, and unreliable communication links, leading to inaccurate data readings. Reliable data storage is demanding now-a-days for storing data on local monitoring station as well as in online live server through internet. In current solutions generally record is being saved on the database timestamp. Sometime internet is not working properly due to congestion and there is frequent packet loss. Current solutions often synchronize records based on database timestamps, leading to record duplication and waste storage. Both databases synchronize each other after internet restoration. By providing synchronization among nodes and data, accuracy and storage will be saved in IoT based WSNs for precision agriculture applications. A prototype Node-MCU internal memory is used as a resource for achieving data synchronization. This proposed work generates record ID from Node MCU EEPROM which helps in records synchronization if there is any packet loss at the local server or at the online server to maintain synchronization accuracy despite unreliable communication links. Experiment shows that for a particular duration Node MCU generated 2364 packets and packet loss at local server was 08 and at online server was 174 packets.  Results shows that after synchronization 99.87\% packets were synchronized. Using previous technique of timestamp, the redundancy was 70\% which reduced to 0\% using our proposed technique. 
\end{abstract}

\maketitle

\section{Introduction}
\subsection*{Overview}
  Wireless Sensor Networks (WSNs) have evolved into a foundational technology across diverse sectors, facilitating the immediate surveillance and acquisition of data for a broad spectrum of purposes. In contemporary times, precision agriculture has surfaced as a pivotal field of application, utilizing smart systems to enhance crop yield through the observation and regulation of environmental factors. Wireless Sensor Networks play a vital role in precision agriculture by collecting data from distributed sensors deployed in the field, including soil moisture, temperature, humidity, and light intensity. However, ensuring accurate and synchronized data collection from numerous sensor nodes poses a significant challenge in WSNs. Timing issues arise when attempting to synchronize the timestamps of two types of data that are obtained from distinct sensor nodes \cite{tamai2024wireless}.
  
The emergence of WSN as a potent technology in precision agriculture applications facilitates effective oversight and management of agricultural processes, ensuring streamlined monitoring and control. The progression of the automation domain has been propelled by IoT networks comprising sensors and actuators. In this framework, distinct application areas establish distinct criteria encompassing transmission medium, range, throughput, communication jitter, and more \cite{romanov2023enabling}. In precision agriculture, accurate and synchronized data collection is crucial for effective analysis and decision-making.

While selecting the routing protocol, multiple design conditions to be adhered. This would depend on the node, type of delivery data model in the network, techniques for scalability and adaptability of node deployment, the operating environment, device type, and the nature of migration \cite{karthick2023energy}. Traditional synchronization methods, such as Global Positioning System (GPS)-based time synchronization, may not be suitable for precision agriculture due to their limitations in terms of cost, coverage, and accuracy in remote agricultural areas.

\subsection*{Synchronization}
\begin{figure}[]
    \centering
    \includegraphics[width=12cm]{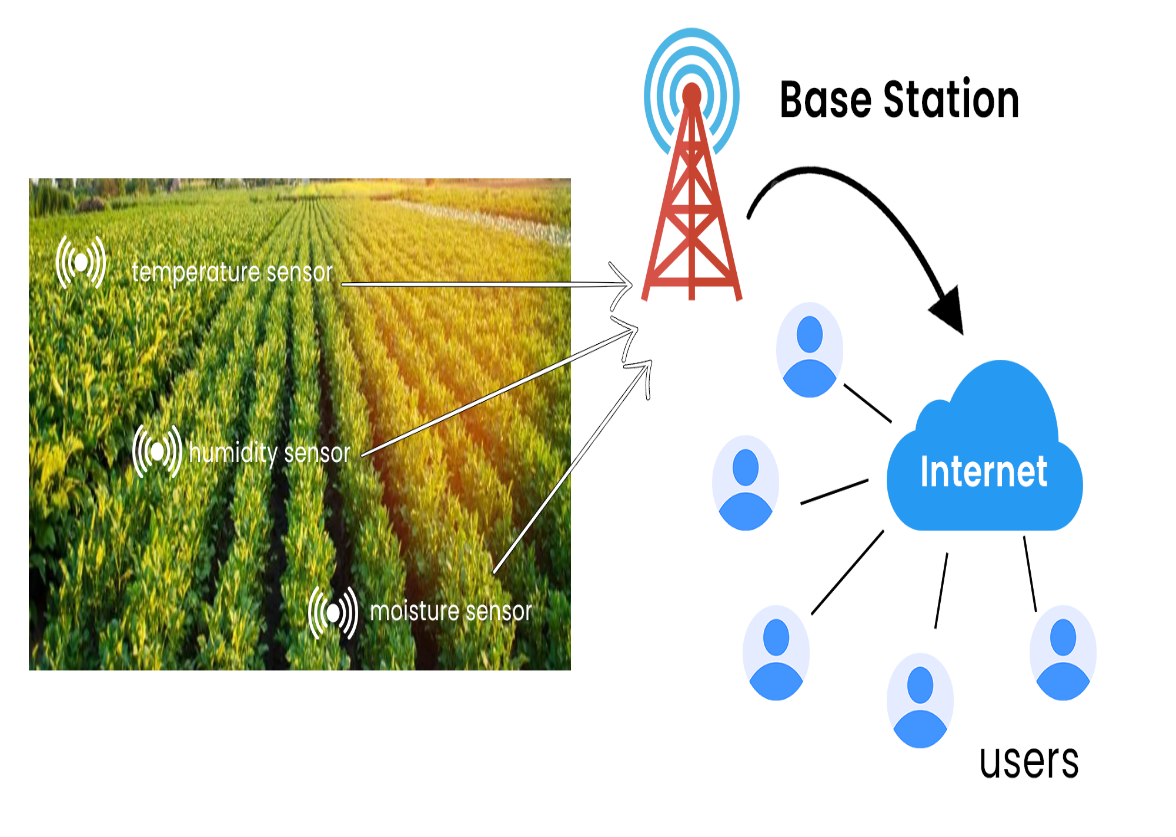}
    \caption{Data collection using WSN}
    \label{fig:Fig1}
\end{figure}

Synchronization is crucial for WSNs to assure that the collected data represent the existing state of the environment accurately. In precision agriculture applications, where timely and precise information is critical, achieving synchronization is of utmost importance. Data synchronization in WSNs refers to the process of aligning the clocks of individual sensor nodes to a common time reference as shown in Figure \ref{fig:Fig1}. It plays a pivotal role in comprehending the temporal correlations among various environmental variables, thereby enabling well-informed choices pertaining to irrigation, fertilization, pest management, and other agricultural methodologies. Within the realm of precision agriculture, synchronized data collection holds notable significance. There are various   challenges in   the industry in agriculture field that is dependent on information technology \cite{kasturi2023iot}.
\subsection*{Adaptive synchronization for precision agriculture WSNs}
In WSNs, synchronization is necessary for precise data analysis. It accomplishes accurate measurements, makes coordination of actions easier, and guarantees smooth data transfer between devices \cite{masalskyi2024synchronization}. A detailed analysis of the proposed synchronization approach is represented; highlighting its benefits and performance characteristics compared to existing synchronization techniques. The system was evaluated through extensive simulations and real-world experiments, demonstrating its effectiveness in achieving accurate and timely data synchronization in precision agriculture WSN. Furthermore, the proposed synchronization mechanism adapted dynamically to changes in the network topology and optimizes energy consumption, making it suitable for resource-constrained WSN deployed in precision agriculture applications. The algorithm adjusted the synchronization intervals based on the network conditions, minimizing unnecessary synchronization cycles and conserving energy resources. This adaptive behavior ensures efficient data collection and synchronization while maintaining the desired precision.

On the other hand, a WSN might behave like an attractive free alternative to wired systems because of its ease in deployment for far-off places and they can finally also be considered a self-organizing platform. The proposed algorithm addresses the challenges associated with node failures and communication disruptions, ensuring continuous synchronization and reliable data collection in dynamic agriculture. By achieving accurate and synchronized data collection, farmers and agronomists can make informed decisions based on real-time information. This can lead to optimized resource allocation, timely interventions in case of crop diseases or pests, and improved overall agricultural productivity. The affordability of our method renders it appropriate not only for extensive agricultural enterprises but also for modest-scale farmers. By providing synchronization among nodes and data, accuracy and storage will be saved in IoT based WSNs for precision agriculture applications. The suggested approach holds the potential to substantially amplify the dependability and efficiency of precision agriculture systems, empowering farmers and agronomists to make well-judged choices grounded in precise and harmonized sensor data.

\subsection*{WSNs in precision agriculture applications}
WSNs are revolutionizing precision agriculture by offering detailed, real-time data on diverse environmental and crop parameters. These networks deploy numerous interconnected sensors across agricultural fields, monitoring critical factors such as soil moisture, temperature, humidity, and crop health. This precision conserves resources, reduces waste, and enhances plant growth, leading to higher crop yields and more sustainable farming practices. The integration of WSNs with advanced data analytics and machine learning algorithms further optimizes decision making processes, enabling predictive maintenance and proactive management. Multiples nodes are connected to a sink node so that they can collect data. After collecting data, they can use that data for processing and analysis. However, due to limited communication range, some nodes cannot directly connect with the sink \cite{liu2023routing}. As a result, WSNs not only improve productivity and efficiency but also contribute to the environmental sustainability of agriculture by reducing the reliance on chemical inputs and minimizing the ecological footprint.

WSNs also play a vital role in climate control by monitoring weather conditions, to enable farmers to adjust their practices in response to changing environmental factors, thus reducing the risk of crop damage due to extreme weather. Furthermore, this includes some of the verticals such as irrigation, monitoring soil moisture, fertilizer optimization and control, and energy conservation which goes with food cultivation across any sort of agricultural domains therefore requires explicit integration among IoT and WSN \cite{mowla2023internet}. In agriculture field, wireless sensor networks have many hopeful applications such as gathering the farm's soil moisture, temperature and humidity data collected by distributed sensor over a region of interest in addition to light intensity Overall, the integration of Wireless Sensor Networks in precision agriculture leads to improved efficiency, reduced environmental impact, and higher productivity by providing actionable insights and enabling precise control over farming practices.

\subsection*{Paper Structure}
The research paper is organized as follows: Section 2 surveys some of the related articles to ground the research. Section 3 describes the detailed approach and the proposed methodology, along with steps and techniques involved in carrying out those steps. Section 4 includes the experimental setup and the prediction results in detail, discussing data and outcomes. This is followed by the conclusion at the end, summarizing the key findings and implications of the study.

\section{Literature review}
The literature review focuses on WSN data synchronization using Node-MCU memory for precision agriculture applications. It explores existing research and studies related to the synchronization of data in wireless sensor networks (WSNs) specifically using Node-MCU memory as a synchronization mechanism. The review aims to depict the existing state of knowledge, research gaps, and potential options in leveraging Node-MCU memory for achieving accurate and reliable data synchronization in the context of precision agriculture applications. By analyzing the previous researches, this analysis provides a comprehensive understanding of the advantages, limitations, and future directions of utilizing Node-MCU memory for WSN data synchronization in precision agriculture.

Dependable tools for agricultural management are provided by wireless sensor networks. A cost-effective and energy-efficient wireless sensing system tailored for agricultural settings was introduced \cite{tagarakis2021low}. Geared towards small to medium-sized fields, this system incorporates a pseudo device functioning as a Coordinator. This Coordinator processes and transmits data packets to the cloud via a 4G cellular network. Powered by solar energy, the system functions autonomously, exhibiting commendable performance in terms of reliability, communication range, energy self-sufficiency, and uninterrupted operation. This technology holds the potential to effectively monitor environmental, soil, and crop parameters.

New technological development has led to the advancement of precision agriculture systems, which offer significant benefits in cost reduction and energy conservation for the agricultural sector. In a study \cite{rodriguez2020autonomous}, a prototype precision agriculture system tailored for small and medium-sized plantations was introduced. This system integrates Wireless Sensor Networks (WSNs), sensor devices, Internet of Things (IoT) technologies, and data analysis to improve agricultural practices. The proposed system focuses on optimizing energy management and maintaining affordability while enhancing the efficiency of farming operations. It collects environmental data through a network of sensors, which is then analyzed to make informed decisions about irrigation. The system activates the watering system based on real-time data, ensuring that water is used efficiently and only when needed. The study provides a comprehensive analysis of the agricultural monitoring system, detailing its design, implementation, and the outcomes achieved. By combining these technologies, the system aims to offer practical solutions that enhance agricultural productivity while minimizing resource usage and operational costs.

Accurate time synchronization is crucial for various functions like data fusion in Wireless sensor network mainly in the decentralized systems, TDMA schedules, and synchronized sleep periods. Previous time synchronization techniques were not initially designed for wireless sensor networks and thus need extension or redesign. Time synchronization technique was designed for Wireless sensor networks \cite{sichitiu2003simple}. Minimal complexity was suggested by proposed method to achieve high accuracy in terms of network Bandwidth, storage and processing. It is highly suitable for sensor networks, offering tight, deterministic bounds on both offsets and clock drifts. A technique was proposed to synchronize the whole network for the preparation of data fusion with the execution of proposed methodology evaluated based on real world experiment of Wireless ad-hock network in the composition of data fusion.

Full time stamp is required in Wireless sensor networks in previous existing clock synchronization approaches to accurately determine clock skew and offset. However, in many scenarios aimed at reducing energy consumption, minimizing communication overhead, or enhancing security, timestamps are not swapped during the coordination process. This leads to nodes lacking full time information, presenting a significant challenge in jointly estimating clock skew and offset with limited observations. A Partial Timestamp synchronization scheme was proposed in which they used only local timestamps and the relationships obtained from basic synchronization protocol as addressed in \cite{wang2023clock}. When stochastic delays followed a Gaussian distribution based on PTS, a 
joint Maximum Likelihood Estimator was established to estimate clock skew and offset. Additionally, they present a Best Linear Unbiased 
Estimator for PTS with exponential random delays, capable of simultaneously estimating clock skew and offset. Finally, simulations demonstrate the effectiveness of the proposed algorithms.

Experimental research often requires recording rapid processes that are difficult to localize precisely in space and time, necessitating a network of recording devices to capture critical events and gather sufficient data. Synchronizing data from these spatially distributed devices is crucial, as discrepancies in timekeeping can distort the understanding of the studied process. A solution was proposed by synchronizing the local clocks of wireless nodes with an accuracy of  \SI{12}{\micro\second}/second using low-cost commercial components \cite{zuev2023wireless}. This method, tested with ESP family microprocessors, offers a practical and cost-effective approach. It has been applied to study high-velocity injuries using artificial body simulators and is versatile enough for measuring vibrations in electrical machines, engines, and structural health monitoring, demonstrating its broad utility and effectiveness in various fields.

The reason why localization within wireless sensor networks has received quite a lot of attention from analysts is because it has many real time applications that help identify the origin of events. Wireless sensors also transmit multimedia data such as audio, video and images in addition to scalar data. These WSNs are therefore known as WMSN. Consequently, various techniques for sensing node identification in wireless sensor network have been developed by researchers through simulating real-time situations or using simulations. In the article \cite{wajgi2024localization}, clustering-based localization strategies for WSN (Wireless Sensor Network) and Wireless Multimedia Sensor Networks) was discussed in depth. Clustering is one of the ways used in wireless sensor networks and wireless multimedia sensor networks to provide distributed computing and parallel processing with each group operating independently with its own small subgroups. Clustering techniques would be highly useful to improve traditional localization procedures’ performance. By dividing the network into partitions that can operate independently thus reducing communication cost, clustering reduces communication overheads significantly. Additionally, distributed computing improves throughput of a network through clustering. The paper also raises future concerns and issues so as to know how new researchers should go about future studies on WSN.

Precision agriculture is increasingly essential for feeding the growing global population, minimizing environmental impact, and optimizing scarce resources. However, the high cost of technology adoption poses a significant challenge. This paper addresses this issue by introducing a low-cost wireless sensor network comprising soil moisture sensors. These sensors, built with coils wound on plastic pipes, enable moisture measurement at different depths. Through mutual induction, the sensors monitor soil water content, aiding farmers in optimizing irrigation processes. Several prototypes underwent testing, and the most effective among them showcased winding ratios of 1:2, 15, and 30 spires, functioning at a frequency of 93 kHz \cite{lloret2021wireless}. A specialized communication protocol was devised to bolster the overall performance of the system. A experiment was performed on citrus tree area, in which wireless network was assessed and some of the evaluations including the coverage and received signal strength indication to measure potential signal losses just because of vegetation.

In another study, it was presented that wireless sensor networks are widely used in fields like engineering, science, agriculture, surveillance, military applications, and smart cars \cite{thakur2019applicability}. Precision agriculture extensively adopts WSNs to measure environmental parameters, enhance crop quality and quantity, and conserve natural resources. This review identifies WSN technologies for precision agriculture, analyzes their impact, and explores various environmental parameters, crops, communication technologies, and sensors used. Several research questions are addressed to understand the influence of WSNs in agriculture.

Time synchronization poses challenges in wireless sensor networks, due to clock deviations and the need to associate sensor measurements with reference clock time \cite{skiadopoulos2019synchronization}. In IoT environments, limited capabilities, energy constraints, robustness, and 
extreme conditions (such as deployment in soil) prompt researchers to explore synchronization schemes that relax requirements. This article presents a lightweight synchronization algorithm tailored for wireless sensor networks. The algorithm primarily concentrates on synchronizing measurements on a per-hop basis within the data packets sent to the sink node. The core objective is to achieve synchronization of data rather than synchronization of node clocks. This approach minimizes unnecessary overhead by eliminating the requirement for supplementary messages and re-synchronization intervals. The analysis of the proposed algorithm illustrates its reliance on residual time, distance, and average skew deviation. Through simulations, the algorithm's effectiveness is thoroughly exhibited, and analytical results are validated. A comparison with a traditional synchronization algorithm shows that the proposed approach is suitable for demanding IoT systems.

Efficient data collection is vital for precision agriculture and smart agriculture, which rely on wireless sensor networks (WSNs). However, agricultural WSNs face challenges like multitasking, data quality, and latency. To address these issues an innovative solution using edge computing-enabled Wireless sensor networks was proposed \cite{li2020edge}. They introduced a novel data collection framework that integrates Wireless sensor networks and edge computing, considering multiple sensors and tasks. Utilizing a double selection strategy, the edge computing server identifies the optimal node and sensor network by considering task demands and sensor interconnections. Furthermore, an algorithm for data collection is developed to enhance data quality through predefined values optimization. Through simulations, the proposed strategy is evaluated, showing superior performance compared to traditional methods across various metrics.

Within precision agriculture, Wireless Sensor Networks (WSNs) play a pivotal role by enabling the effective utilization of natural resources through real-time data collection on farm premises. This data empowers farmers with the knowledge needed to make well-informed and intelligent choices, ultimately leading to improved crop yields and enhanced profitability. Another work introduced an innovative method for monitoring agricultural lands through the utilization of a Wireless Sensor and Actuator Network (WSAN) \cite{tiglao2020agrinex}. The Agrinex system employs a mesh-like arrangement of in-field nodes that function both as sensors, measuring parameters like soil moisture, temperature, and humidity, and as actuators, granting control over drip irrigation valves to enhance water management efficiency. The network is designed to dynamically self-reorganize when changes occur. The Agrinex system demonstrates great potential as a WSAN framework for various agricultural applications.

The data collection though Wireless sensor is the basic source of energy dissipation in a low-cost wireless sensor network. Motivated by this, compressive data collection has become an important tool for minimizing energy consumption. Compressive data collection. Compressive sensing was used for minimizing the amount of transmitted data to reduce the energy dissipation. Keeping in view, physical interference model was used. They are jointly in a disconnected network and it investigates the compressive sensing-based data collection and link scheduling. With the network disconnected, mobile collets have been developed to enable data collection within the network. Compressive sensing is when active sensors are triggered to send compressed data that mobile collec tors then recover against all the sensed data. Another study targeted two metrics, end-to-end latency and number of transmissions for data collection with this work \cite{ghosh2023energy}. Since the joint problem is NP-Hard, researchers have tried to find heuristic-based solutions for tree construction as well as link scheduling. The feasibility of the proposed algorithm is validated by simulation results in comparison with some existing algorithms.

Over the past few decades, Wireless Sensor Network (WSN) technology has gained extensive acceptance across diverse scientific domains, enabling accurate monitoring of climate phenomena such as air pollution, as well as catastrophic incidents like landslides. Its applications in agriculture for field monitoring have been particularly notable. WSN represents an emerging technology that has proven to be a valuable tool in exploring the intricate world behind environmental phenomena using 
small sensor nodes. "Expert Advisory System" developed to enhance farmers' productivity, save time, and improve crop efficiency \cite{muzafarov2019wireless}. The system utilizes WSN for real-time monitoring of crop fields and delivers essential information to farmers through the Internet. Based on the received information, farmers can take necessary remedial actions. The study also includes the simulation of WSN using the Contiki Simulator tool and considers queuing models for WSN in the research.

Another technique used to reduce the mass and design complexity of cable harnesses, the space industry seeks to deploy reliable Wireless Sensor Networks (WSNs) in spacecraft and launcher systems \cite{lubken2024adaptive}. Low-latency sensor data acquisition in these systems depends on synchronized local clocks across network nodes, making time synchronization essential. Controlling the local clock on a node between synchronization points provided by a time distribution protocol is challenging, especially in harsh environmental conditions like temperature fluctuations and electromagnetic interference, which cause rapid clock drift. Existing methods struggle to mitigate these large drift rates and cope with varying drift amounts. This article introduces the Adaptive Local Clock Control (ALCC) algorithm, which minimizes deviation from a reference clock under extreme conditions, outperforming existing methods by nearly an order of magnitude. A formal model of Adaptive Local Clock Control is compared to current solutions in simulations replicating real-world scenarios, and the algorithm is verified on a hardware test platform, demonstrating its effectiveness in maintaining synchronized local clocks under harsh conditions.

In the Industrial Internet of Things (IoT), energy efficiency is critical for operational longevity, and traditional approaches like flooding for time synchronization often waste energy through redundant message transmissions. Intelligent neighbor-knowledge synchronization (INKS) method was introduced to address this issue by leveraging each node's knowledge of its neighbors to optimize the synchronization process, thereby reducing the total number of synchronization messages and conserving energy \cite{jia2024low}. Implemented and evaluated using real wireless sensor networks with various configurations, intelligent neighbor-knowledge synchronization demonstrates superior performance to existing techniques such as rapid flooding multiple one-way broadcast time synchronization (RMTS). Simulations on large-scale networks, particularly with a four-way grid topology, reveal that intelligent neighbor-knowledge synchronization reduces transmitted messages by approximately 72\% compared to RMTS and matches the efficiency of scheduling-based low-energy synchronization for IoT, highlighting its potential for enhancing energy efficiency in IoT applications.

The elucidation of agricultural automation presented \cite{shafi2019precision} via IoT has the potential to revolutionize the sector, imbuing it with dynamism and efficiency. At the heart of this transformation lie Precision Agriculture and Wireless Sensor Networks. The study in question significantly contributes by meticulously examining how sensor networks and remote sensing technologies synergistically enable real-time surveillance of crop well-being and environmental conditions within agriculture. Furthermore, it proposes an IoT-driven solution that harmonizes wireless sensors and remote sensing platforms to accomplish effective crop health monitoring. Additionally, the research incorporates a case study demonstration, addresses the existing challenges, and outlines the future prospects within this domain.

A cost-effective and energy-efficient solution is proposed for information monitoring in large-scale agricultural farms. The system utilized sensor-based soil properties measurement to automate farm operations and improve results compared to manual methods \cite{saqib2020low}. The paper presented an innovative low-power, long-range communication module with a tree-based mechanism for efficient agricultural monitoring. Sensor nodes equipped with essential components enable successful communication up to 750 meters with minimal packet loss, making it a valuable reference model for large-scale agricultural applications.

In another research, authors proposed an innovative architecture designed to enhance security protocols within wireless sensor networks (WSNs) and Internet of Things (IoT) frameworks \cite{hasan2023efficient}. Their study addresses 
several critical aspects, including energy consumption, mobility, information transmission, quality of service (QoS), and overall security. By 
focusing on the data link layer of WSN frameworks, the research tackles existing challenges, limitations, and protocol issues that are often overlooked in the broader literature on WSNs. The authors highlight that, despite extensive research on WSNs, there is limited exploration of how the data link layer impacts network performance. Their work emphasizes practical solutions that incorporate machine learning to improve various aspects of WSN performance. Key challenges identified in their study include managing flow control, ensuring QoS, maintaining security, and optimizing overall performance in WSN applications. The proposed architecture aims to address these challenges by integrating advanced security protocols and optimizing network parameters, thereby enhancing the efficiency and robustness of WSN and IoT systems.

In another study, domain-agnostic monitoring and control framework (MCF) was introduced for IoT systems which is built on building blocks that can be implemented by the five-layer architecture of IoT system and includes three sub-domain subsystems: Monitoring, Controlling, Computing \cite{senoo2023monitoring}. They have seenn it successful in smart agriculture using commodity sensors and code, and an affordable technique. Our MCF offers scalable, reusable, and interoperable solutions, costing up to 20 times less than commercial alternatives. It operates efficiently on rechargeable batteries and solar power, with stable data exchange and minimal power consumption. The framework supports reliable data from multiple sensors, ensuring accurate and consistent performance over time.

Extensive blueprint and execution of an economical Wireless Sensor Network (WSN) system designed specifically for intelligent agriculture, with a focus on synchronization \cite{zervopoulos2020wireless}. The study's objective was to create a cost-effective and proficient answer for real-time surveillance within agricultural environments. Through utilizing the sink node's clock as a reference point, a straightforward yet impactful synchronization mechanism was put forth. This approach facilitated the harmonious synchronization of data measurements across the entirety of the WSN. The system's performance was rigorously tested in an olive grove, demonstrating successful measurement synchronization even in the presence of slight time differences resulting from transmission delays. The findings underscore the system's capability to provide crucial environmental data in precision agriculture applications, paving the way for more informed decision-making and resource optimization in farming practices. The study's cost-effectiveness and reliability make it a valuable contribution to the advancement of smart agriculture technologies.
   
\subsection*{Research Gap}
Achieving accurate data synchronization and efficient storage in 
Wireless Sensor Networks for precision agriculture applications involves overcoming several complex challenges. Timestamp-based solutions often lead to issues such as record duplication and storage inefficiencies, which compromise the effectiveness of data analysis and decision-making. Advanced synchronization protocols are essential to ensure precise time alignment of sensor data, thereby eliminating redundant records and maintaining data integrity. Additionally, optimized data storage solutions are necessary to handle the large volumes of high-frequency data generated by sensors, which can otherwise lead to inefficiencies and bottlenecks. Internet congestion exacerbates the problem by causing packet loss and incomplete data transmission, further complicating reliable data management. To address these issues, it is crucial to develop and implement sophisticated techniques for data synchronization, duplication prevention, and robust transmission protocols, ensuring that data is accurate, complete, and efficiently managed for improved agricultural productivity and resource optimization.

\subsection*{Problem Statement}
Accurate data synchronization among sensor nodes in Wireless Sensor Networks (WSNs) is crucial for precision agriculture applications, as reliable data storage is now highly demanded for local monitoring stations and online live servers through the internet. In precision agriculture, accurate and synchronized data collection is crucial for effective analysis and decision-making. Current solutions often save records using timestamps, which can lead to record duplication and storage wastage, impacting analysis and decision-making processes. Enhanced synchronization protocols and data duplication removing techniques are needed to ensure accurate time alignment of data from different sensors and to identify and remove redundant records. Additionally, optimized data storage solutions are essential to handle high-frequency data efficiently, thereby improving data quality, analysis accuracy, and decision-making in precision agriculture.

\section{Materials and methods}
\begin{figure}[]
    \centering
    \includegraphics[width=12cm]{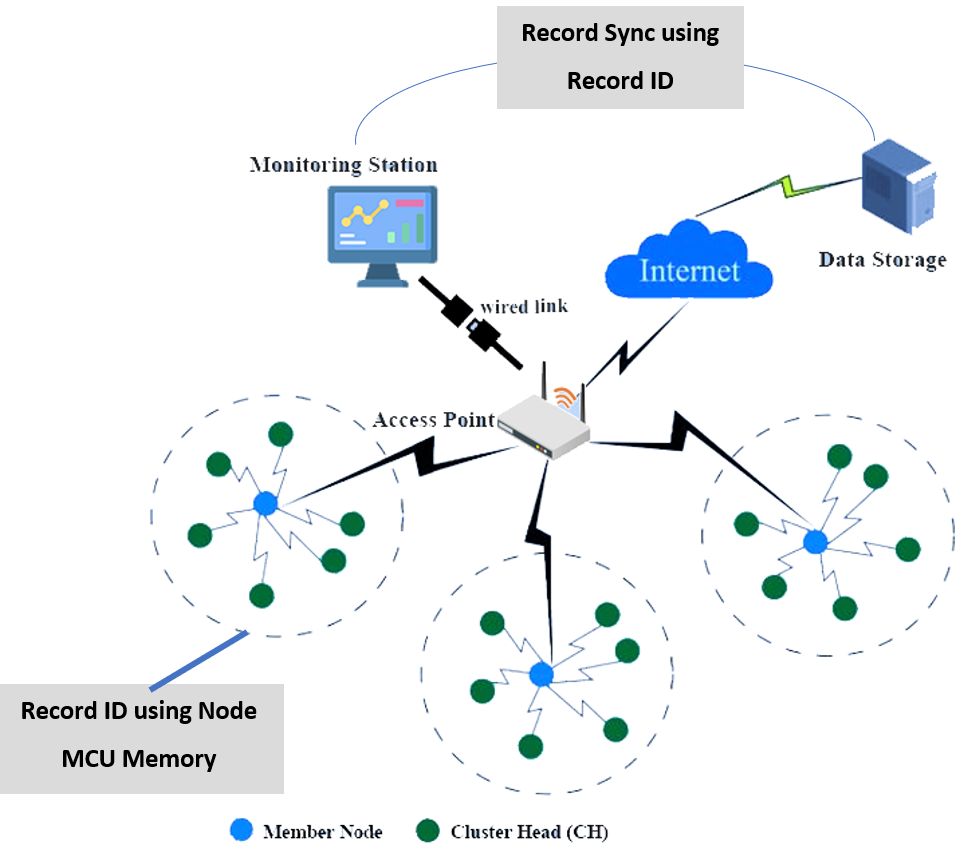}
    \caption{IoT architecture for WSN}
    \label{fig:Fig2}
\end{figure}

The approach suggested employs the memory capabilities of Node-MCU devices to enable the synchronization of data among the sensor nodes within the Wireless Sensor Network (WSN). Figure \ref{fig:Fig2} illustrates a representative IoT architecture for Wireless Sensor Networks. A synchronization algorithm is developed, incorporating a distributed consensus mechanism to ensure robust handling of node failures and maintenance of synchronization accuracy, even when dealing with 
unreliable communication links within the WSN. The effectiveness and reliability of the proposed approach are validated through extensive experiments, testing the synchronization performance under different scenarios including node failures and unreliable communication conditions.

Moreover, within the domain of IoT, there exist further hurdles tied to energy consumption. These challenges include packet losses, collisions, issues in the physical channel, frame overhearing, protocol overhead, and processing overhead. Packet losses and collisions lead to the need for retransmissions, which consume additional energy. In the last couple of years, the field of agriculture has witnessed much research on modern senor techniques.

The way in which intelligent sensor is being used effectively and efficiently in diverse activities have been resulting in optimized resource utilization and minimization of human interruptions \cite{kabir2022environmental}. Frame overhearing occurs when nodes receive and process packets not intended for them, leading to unnecessary energy use. Consequently, the quest for energy-efficient and sustainable IoT solutions has risen to the forefront of research priorities for scholars in this sphere. Researchers are focusing on developing advanced algorithms and protocols that minimize energy use, exploring energy-harvesting technologies, and optimizing network architectures to enhance the longevity and sustainability of IoT devices and systems. 

\subsection*{System architecture design}
The overall system architecture for the WSN-based precision agriculture application involves strategically deploying sensor nodes throughout the agricultural field to ensure comprehensive data coverage. In our real-world experiment, we utilized the Node MCU ESP 8266 model paired with the DHT 22 sensor to measure temperature and humidity. ESP-NOW is a communication protocol designed by Espressif Systems for low powered and inexpensive IoT devices that operate in situ. ESP-NOW supports both way communication between more than one transmitter and many end-devices \cite{magzym2023synchronized}. For the local server, we used XAMPP version 8.2.12 (64-bit), and we developed sensor controller using Arduino IDE version 2.3.2. Our database server was MariaDB version 10.5.20, running PHP version 7.3.33, and we managed the database using phpMyAdmin version 5.2.1.

The Node MCU board acts as the main synchronization resource, utilizing its memory capacity and Wi-Fi capabilities. Sensor nodes, equipped with sensor, are placed at key locations to monitor environmental parameters like temperature and humidity, providing maximum coverage and accurate data representation. It collects data from the DHT22 sensors, stores it temporarily, and transmits it simultaneously to both local and online servers for real-time 
monitoring and analysis. This setup ensures continuous data flow, synchronization accuracy, and robust network performance, essential for precision agriculture applications. 

\subsection*{Sensor node configuration and protocol}
In this setup, sensor nodes are outfitted with appropriate sensors designed to gauge environmental factors such as temperature, humidity, soil moisture, and the developmental stages of crops. The majority of protocols for wireless networks need. The access point, gateway or cellular tower are mostly required for protocols of wireless sensor networks to broadcast a beacon at regular intervals with timestamps. Furthermore, a resulting device is used in order to synchronize its internal clocks \cite{puckett2023ecosync}. Communication protocols are established to streamline the transfer of data from the sensor nodes to the central base station, utilizing the inherent Wi-Fi capabilities of Node-MCU devices. For sending data from the Node MCU to the database, typically, the HTTP protocol is used, especially when interfacing with a web server such as XAMPP. The Node MCU sends HTTP requests to a PHP script on the server, which then processes the data and inserts it into the MariaDB database.

\subsection*{Methodology}
The process of setting up a sensor system with a Node MCU starts 
with the physical connection between the sensor and the Node MCU’s input pins, ensuring that the sensor is properly powered to 
operate. Careful attention is given to the wiring, as incorrect connections can lead to malfunction or damage to the components. The sensor must receive adequate power, typically through the Node MCU's VCC and GND pins, and its data output pin must be connected to one of the Node MCU’s input pins. Once the hardware connections are securely in place, the system is ready to be initialized. Initialization involves configuring the Node MCU to communicate with the sensor, which may require uploading specific firmware or code that instructs the Node MCU on how to interact with the sensor and read its data. This is mainly due to their greater robustness and ability of WSNs can handle work, which taking a transmission efficiency part on the research area for reserving certain system battery power. Most of the WSN applications are interested and supported by data packet-based infrastructure \cite{devasenapathy2023transmission}.

\begin{figure}[]
    \centering
    \includegraphics[width=14cm]{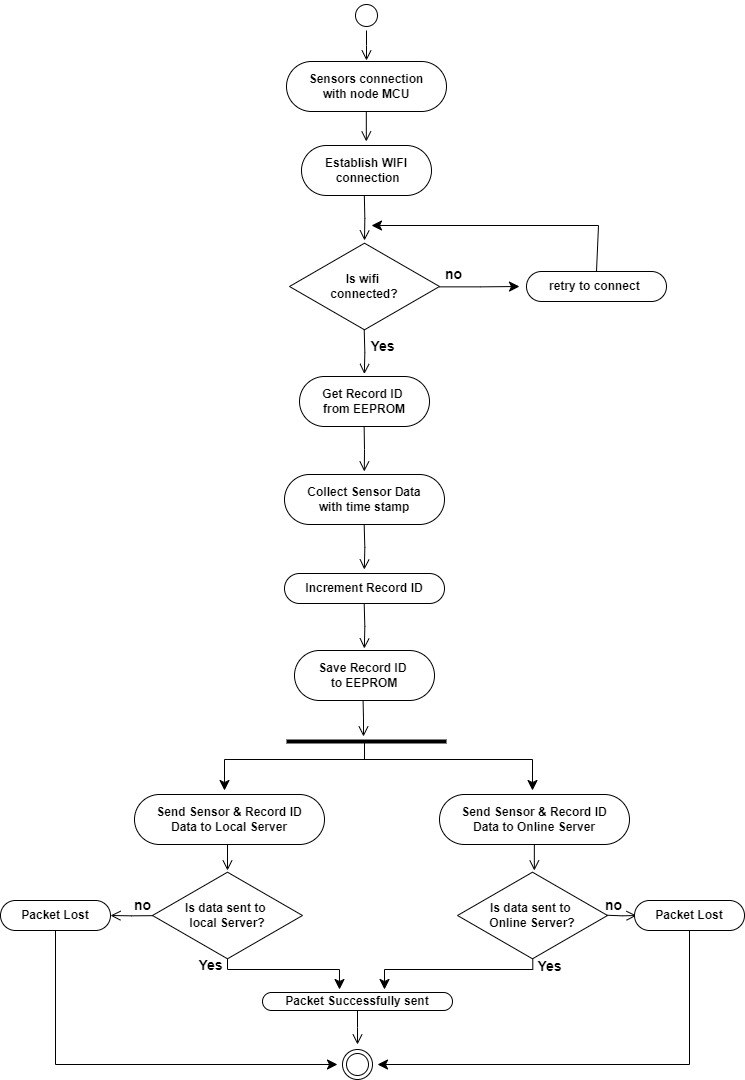}
    \caption{Proposed Methodology}
    \label{fig:Fig3}
\end{figure}

Once the hardware is set up and the firmware is uploaded, the system proceeds to establish a Wi-Fi connection. During this phase, the Node MCU continuously searches for and tries to connect to a specified Wi-Fi network using the provided SSID and password. This process is crucial, as a stable Wi-Fi connection is necessary for transmitting sensor data to a remote server or cloud platform. If the initial Wi-Fi connection attempt fails, the Node MCU is programmed to automatically retry until a stable connection is achieved. Only after a successful connection is established does the system move forward with any data processing or transmission tasks, ensuring that it is fully operational and capable of sending real-time data for further analysis. An activity diagram of the proposed methodology is shown in Figure \ref{fig:Fig3}, illustrating the sequence of steps involved in setting up and initializing the sensor system.

After successfully connecting to the Wi-Fi, the system retrieves a Record ID from the EEPROM (Electrically Erasable Programmable Read-Only Memory). It is a memory chip that keeps data even when the power is turned off. This Record ID is a unique identifier for each set of sensor data collected. The system gathers data from the connected sensor, reading sensor values and recording the exact time each reading is taken, thus associating a timestamp with the data. The Record ID is then incremented by one to ensure that the next set of data has a unique identifier. The updated Record ID is written back to the Electrically Erasable Programmable Read-Only Memory to preserve it for future use.

With the sensor data collected and the Record ID updated, the system prepares to transmit this information. The data, along with the Record ID, is first sent to a local server, which acts as an intermediate storage or processing point. A check is performed to verify that the data was successfully transmitted to the local server; if the transmission fails, the data packet is considered lost, and appropriate error handling procedures are initiated. After that, the data is successfully sent to the local server, the packet is then transmitted to an online server, serving as a central repository for remote access and analysis. Another verification check ensures that the data reached the online server, and if transmissions are successful, the process 
concludes with the confirmation of a successful packet transmission. This system ensures accurate data collection, unique identification, and reliable transmission with mechanisms to handle packet loss and ensure data integrity. 
\subsection*{Synchronization algorithm design}
A synchronization algorithm has been developed to leverage the Node MCU's memory as a key resource for establishing a uniform time reference among sensor nodes. This algorithm focuses on maintaining data integrity and reliability by recovering missing packets on both the local and online servers. 

\begin{figure}[]
    \centering
    \includegraphics[width=12cm]{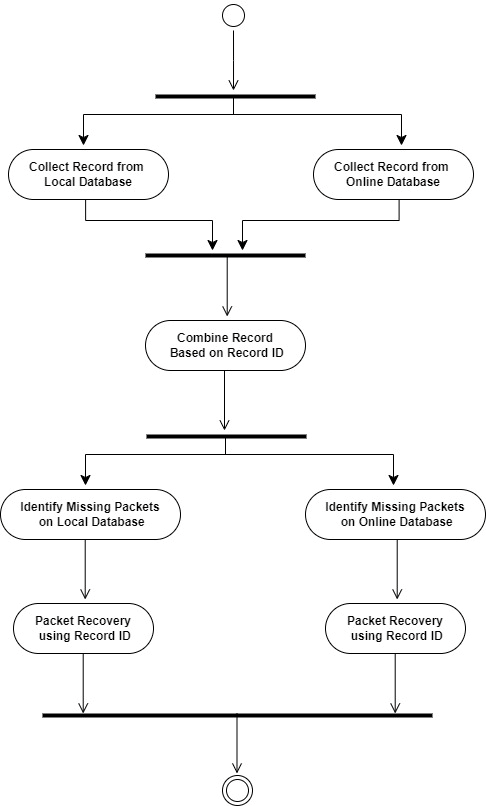}
    \caption{Packet Synchronization}
    \label{fig:Fig4}
\end{figure}

The recovery process uses the Record ID generated by the Node MCU's memory, which is used to store data for every packet. To recover missing packets, the algorithm compares the packets received on the local server with those on the online server. By identifying discrepancies between the two, the system can detect which packets are missing and initiate recovery procedures as shown in Figure \ref{fig:Fig4}.

This approach ensures that all collected data is reliably transmitted and accurately recorded, even in cases where some packets might be lost during the initial transmission phase. This is particularly crucial for the effective operation of Wireless Sensor Network (WSN)-based applications, such as precision agriculture, where the accuracy and consistency of data collection are essential for optimizing agricultural practices and making informed decisions. By establishing a uniform time reference across the network and implementing mechanisms to account for all data packets, the algorithm enhances the overall reliability and robustness of the data collection and transmission process. This approach mitigates the impact of packet loss or delays, ensuring that the complete dataset is preserved and accurately reflects the monitored conditions. Yet a major setback here is the degradation of performance because message delay in synchronization process is non-deterministic which leads to drift and also contributes towards making clock parameter estimations more complex. Consequently, it supports more precise monitoring, analysis, and response, thereby improving the effectiveness of the WSN application in providing valuable insights and facilitating timely interventions.

\subsection*{Data collection using sensor}
Data collection is achieved through a DHT22 sensor connected to a Node MCU module, which measures environmental parameters like temperature and humidity with high accuracy. WSNs not only improve productivity and efficiency but also contribute to the environmental sustainability of agriculture. The Node MCU's onboard EEPROM memory temporarily stores the collected data to ensure that no information is lost in case of temporary disruptions in connectivity or power. After collection, the data is transmitted simultaneously to both a local server and an online server. This dual transmission strategy serves multiple purposes: it ensures immediate local access for real-time monitoring and analysis, and provides remote access and backup through the online server. By having data stored in both locations, the system enhances overall data reliability, accessibility, and redundancy. This approach not only facilitates continuous monitoring but also protects against data loss and enables comprehensive analysis, making the system a robust solution for effective and resilient data management in various applications. 

\subsection*{Node MCU EEPROM memory}
The evaluation of our synchronization approach centered on several critical performance metrics to ensure its effectiveness.  rigorously test our approach, extensive experiments were carried out across various agricultural scenarios, replicating real-world conditions to assess how well the synchronization method performs in dynamic and potentially challenging environments. The anticipated results are expected to demonstrate that our approach not only achieves high synchronization accuracy but also maintains reliable data synchronization even when faced with network fluctuations or node failures.

The contribution of our work lies in offering a practical and cost-effective solution for precise data synchronization in Wireless Sensor Networks (WSNs) used in precision agriculture applications. By leveraging the Node-MCU memory and a specialized synchronization algorithm, our approach provides a scalable method for synchronizing data across numerous sensor nodes. This ensures that large-scale sensor networks can maintain precise and reliable data synchronization, which is crucial for accurate monitoring and decision-making in precision agriculture. The practicality of using readily available hardware, combined with an efficient algorithm, makes this solution both accessible and adaptable, facilitating its implementation in diverse agricultural settings and enhancing overall system performance.

\section{Results and Discussions}
\subsection*{Unique ID in Database}
One possible solution to address data synchronization issues is using a unique ID in the database that auto-increments with each new entry. This ensures that each record has a distinct identifier, preventing duplication. However, when data is transmitted to the database through a Node MCU, a problem arises if the connection is lost. In such cases, once the connection is restored, the Node MCU initialize the record id again with the zero value or if we are using Auto increment in the database, then the database never knows how many records have missed. This can result in gaps or inconsistencies in the data sequence, which may complicate data analysis and decision-making processes.
\subsection*{Implementation}
We implemented our system using the Node MCU ESP 8266 module and sent packets on both a local server and an online server. We sent data to both servers simultaneously, with one set including a record ID and the other set without it. The local server successfully recorded nearly all 
values, missing only a few, while the online server missed many values. This discrepancy was due to issues with the online server and internet congestion. The local server, being on the same network as the Node MCU, had a stable connection, whereas the online server, reliant on internet connectivity, faced challenges like packet loss and network latency, leading to significant data loss. These findings underscore the need for 
robust solutions to ensure reliable data transmission and storage in environments with unreliable internet connectivity, crucial for precision agriculture applications.
\subsubsection*{Packets transmission at online and local server}
\begin{figure}[]
    \centering
    \includegraphics[width=12cm]{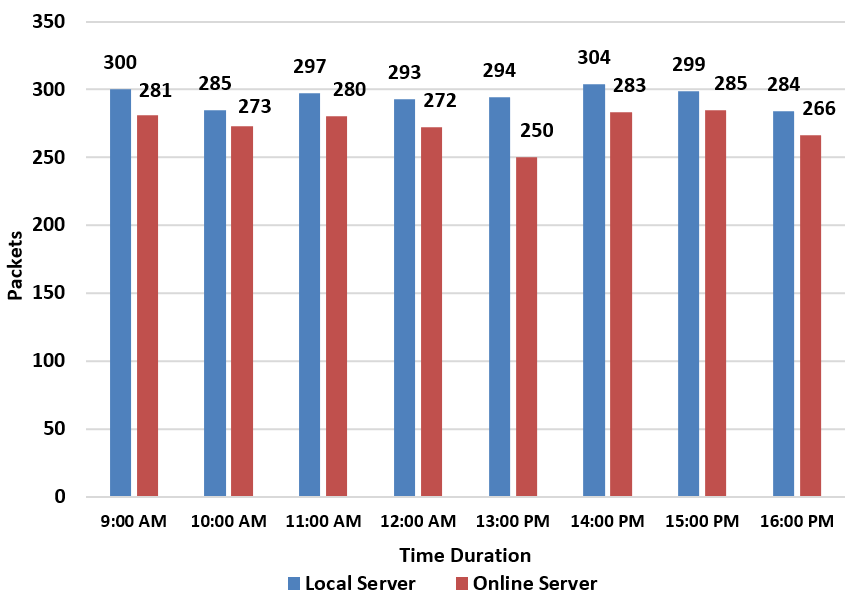}
    \caption{Actual readings of online and local server}
    \label{fig:Fig5}
\end{figure}

Our system was configured to interact with both a local server and 
an online server. The experiment spanned from 08:00 AM to 4:00 PM, covering a total duration of 8 hours. During this period, we recorded a 
total of 2,356 packets on the local server and 2,190 packets on the online server as shown in Figure \ref{fig:Fig5}. This translates to an average of 294 packets per hour on the local server and 274 packets per hour on the online server. Packets were sent to both servers at regular intervals of about 10 seconds. However, this delay increases few seconds due to delayed response from server, particularly the online server. The discrepancy in the number of packets recorded by the two servers could be due to the online server issue or due to internet congestion.
\subsubsection*{Packet loss at online and local server}
\begin{figure}[]
    \centering
    \includegraphics[width=12cm]{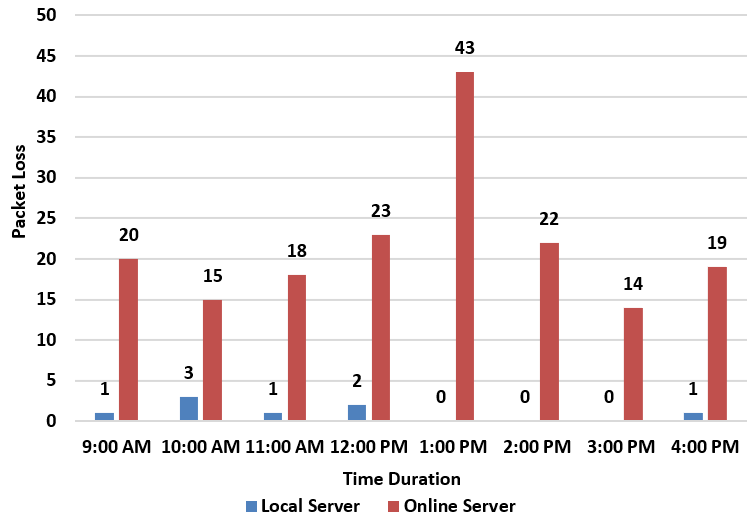}
    \caption{Packet loss at online and local server}
    \label{fig:Fig6}
\end{figure}

After sending packets to the database every 10 seconds over 08 hours span of time. During this process, we noted that 8 packets were lost on the local server. In contrast, the online server experienced a significantly higher loss, with 174 packets failing to be recorded as shown in Figure \ref{fig:Fig6}. To put this into perspective, on average, the online server loses approximately 22 packets per hour. The accuracy and reliability of the data being collected and stored has a directly effect on the rate of packet loss. The higher packet loss on the online server compared to the local server indicates potential issues in the network transmission or server handling that need to be addressed to ensure data integrity. The disparity in packet loss between the two servers highlights the importance of robust error-checking mechanisms and reliable network infrastructure for maintaining data accuracy.

\subsubsection*{Data redundancy of local and online server}
During an 8-hour period of sending data to both the local and online servers, a total of 5,554 packets were transmitted. Upon analyzing the data, we found that 523 packets had identical timestamps, leading to significant data redundancy. This redundancy indicates that the same information was recorded multiple times, which is inefficient. If we synchronize the packets both the online and local servers based on timestamp stored with each record, this results in numerous duplicate entries. These duplicates not only consume additional storage space but also complicate data management and analysis. Efficient storage and accurate data representation are critical, and this level of redundancy can hinder both. Addressing this issue is essential to optimize storage utilization and maintain the integrity and reliability of the data collected. Data redundancy graph of local and online server are shown below in Figure \ref{fig:Fig7} and \ref{fig:Fig8}.   
\begin{figure}[]
    \centering
    \includegraphics[width=12cm]{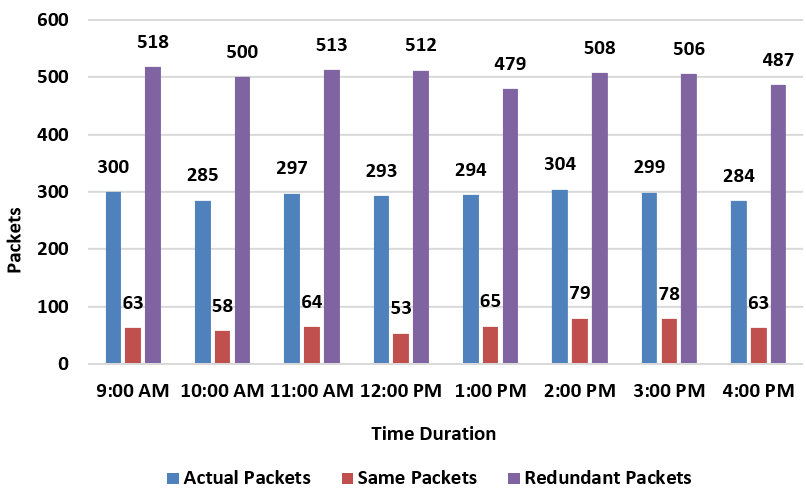}
    \caption{Data redundancy of local server}
    \label{fig:Fig7}
\end{figure}
\begin{figure}[]
    \centering
    \includegraphics[width=12cm]{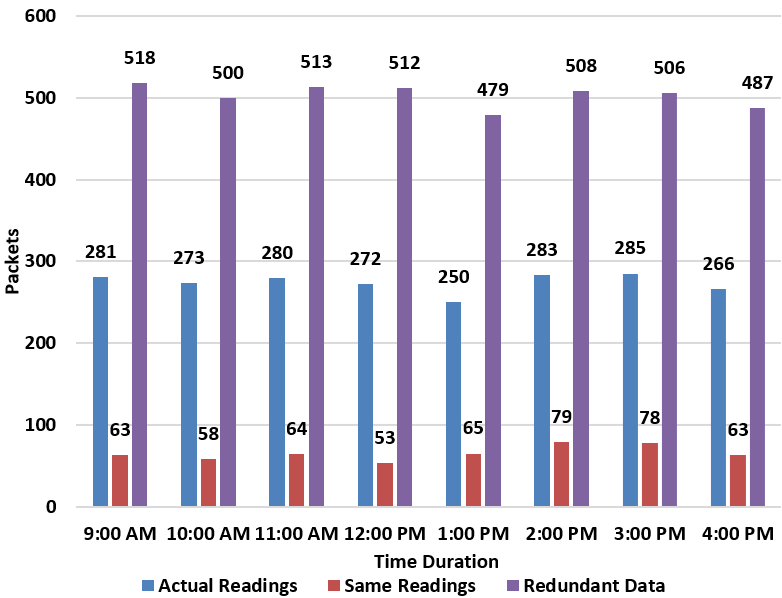}
    \caption{Data redundancy of online server}
    \label{fig:Fig8}
\end{figure}

\subsection*{Record ID packet transmission}
As a potential solution to improve data handling and reduce redundancy, we implemented a system where data values are sent to both the online server and the local server with a unique record ID generated by the Node MCU EEPROM. This unique ID system ensures that each data packet is identifiable and traceable, facilitating synchronization between the two servers. The Node MCU assigns a unique record ID to each data packet before transmitting it, storing the current ID in its EEPROM to maintain continuity. This approach enables the system to resume from the 
last generated ID in the event of a failure or reboot, ensuring that no data 
is lost and that the integrity of the data stream is preserved. 

\begin{figure}[]
    \centering
    \includegraphics[width=12cm]{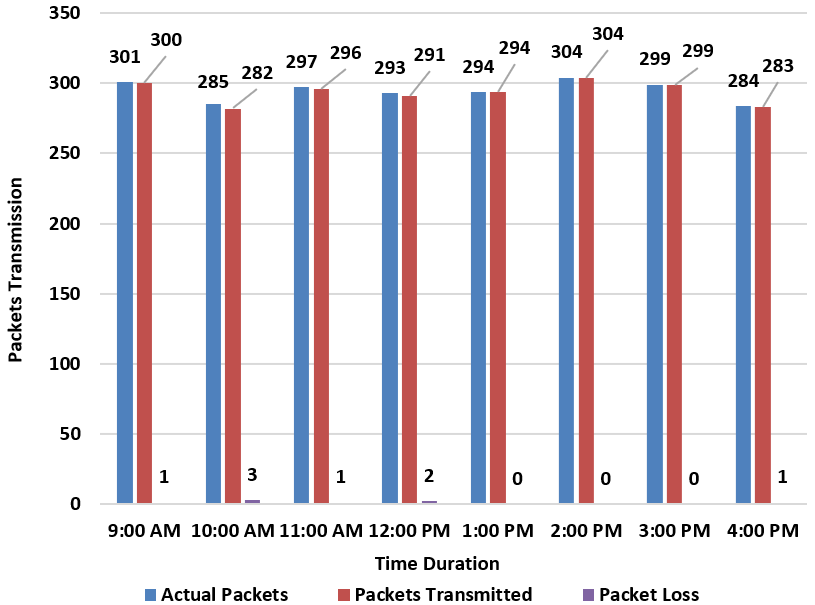}
    \caption{Actual packet transmission using record ID in local server}
    \label{fig:Fig9}
\end{figure}

\begin{figure}[]
    \centering
    \includegraphics[width=12cm]{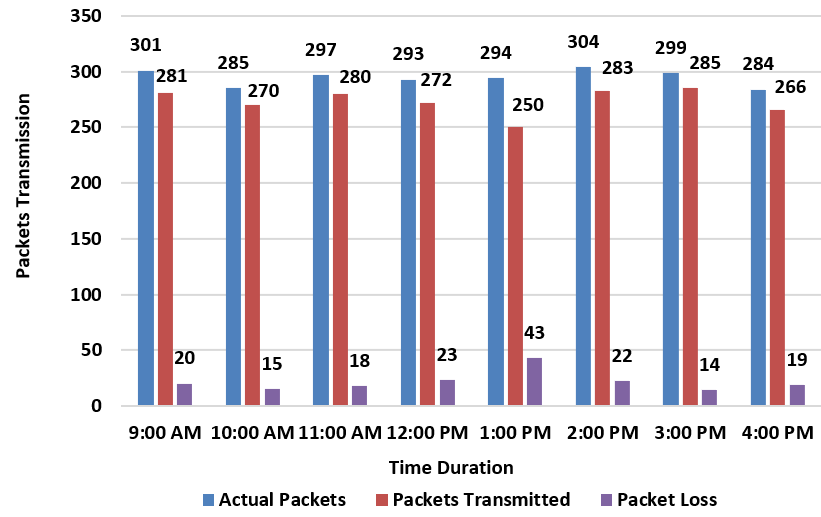}
    \caption{Actual packet transmission using record ID in online server}
    \label{fig:Fig10}
\end{figure}

During an 8-hour period, the Node MCU generated record IDs up to 2364, demonstrating its capacity to handle a substantial volume of data transmissions. The local server successfully recorded 2356 packets, as illustrated in Figure \ref{fig:Fig9}, indicating a high reliability of data reception at the local level. In contrast, the online server recorded 2190 packets, as shown in Figure \ref{fig:Fig10}, revealing a discrepancy that highlights some data loss during transmission to the online server. However, the use of unique record IDs aids in tracking and identifying missing packets, allowing for potential retransmission and recovery of lost data. By comparing the record IDs from both servers, we can pinpoint the exact packets that were lost and take corrective measures to ensure data completeness and accuracy.

The key benefit of using the Node MCU Electrically Erasable Programmable Read-Only Memory is its ability to generate a unique record ID for each packet. This method not only improves data integrity by ensuring that each packet is accounted for but also aids in pinpointing where data loss occurs, allowing for targeted troubleshooting. Additionally, by reducing redundancy and efficiently managing storage, this approach enhances the overall reliability and efficiency of data transmission and storage. Comparing the actual packet transmission of the local and online servers against the generated record IDs reveals that 8 packets were lost on the local server and 174 packets were lost on the online server as shown in Figure \ref{fig:Fig9} and Figure \ref{fig:Fig10}, leading to data inaccuracy.

\subsubsection*{Data synchronization between local and online server}
Data collected from a DHT22 sensor is sent to both a local server and an online server via a Node MCU, with the Node MCU's memory efficiently managing record IDs for each dataset. This ensures that each data packet is uniquely identifiable, aiding in the accurate tracking and synchronization of data. In scenarios where internet connectivity is interrupted, data is transmitted solely to the local server, leading to potential gaps in data transmission to the online server. Despite this, the unique record IDs assigned by the Node MCU facilitate easy identification and retrieval of missing data once connectivity is restored.

Post-collection, a synchronization process is initiated between the local and online servers to ensure any missing data is retrieved and updated on the online server. The Node MCU's unique record numbering system plays a crucial role in this synchronization, enabling precise matching and updating of datasets across both servers. By guaranteeing that all data is accounted for and accurately synchronized, this system supports informed decision-making and enhances the efficiency and effectiveness of agricultural practices.

\subsubsection*{Redundancy comparison with proposed method}
\begin{figure}[]
    \centering
\includegraphics[width=12cm]{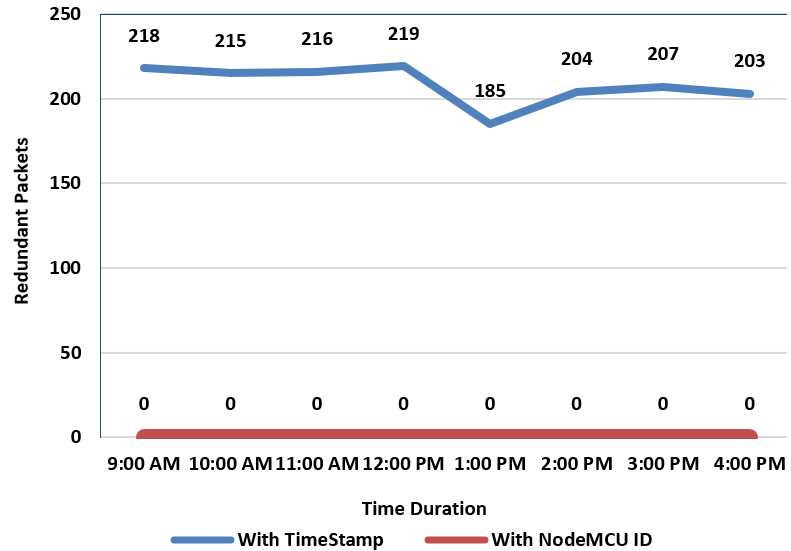}
    \caption{Redundancy comparison with proposed method}
    \label{fig:Fig11}
\end{figure}

After attaching a unique record ID to every packet, the data was saved by both the local and online servers, with only a few packets being missed. When we compared the records using timestamps, we discovered that 1,667 packets were redundant, as illustrated in the Figure \ref{fig:Fig11}. This redundancy indicates that the same packets were recorded multiple times, leading to inefficient use of storage space and potential complications in data analysis.
However, in our proposed technique, where each packet is tagged with a Node MCU-generated ID, we observed no redundant data upon comparison. The Node MCU ID ensures that each packet is uniquely identified, preventing duplicate entries and enhancing the accuracy of the data. By eliminating redundancy, this method optimizes storage usage, allowing for more efficient data management. The benefits of using Node MCU-generated ID are significant. They not only improve data accuracy but also reduce storage costs by preventing the accumulation of duplicate data. This streamlined approach ensures that each piece of data is recorded only once, maintaining the integrity of the dataset and making it easier to analyze and utilize the information effectively.
\subsubsection*{Packets recovery with proposed method}
\begin{figure}[]
    \centering
\includegraphics[width=12cm]{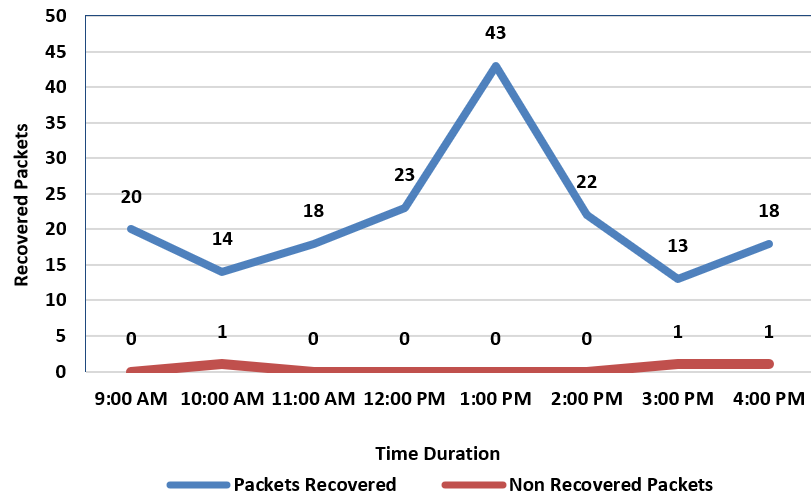}
    \caption{Packets recovery with proposed method}
    \label{fig:Fig12}
\end{figure}

In our proposed method, we successfully recovered lost packets from both servers. The Node MCU generated a maximum ID of 2364, but the local server recorded 2356 packets, indicating a loss of 8 packets. On the other hand, the online server recorded only 2190 packets, resulting in a loss of 174 packets, which is highly considerable. Using our proposed technique, which involves comparing records by their unique Node MCU-generated ID, we were able to recover missing packets from both the local and online servers. Despite the Node MCU generating ID for each packet we observed that only 3 packets could not send to both servers. Since these are not received at either server so could not recovered. The hourly breakdown of recovered packets is illustrated in the Figure \ref{fig:Fig12}. 

This method not only enhances data integrity by recovering lost packets but also provides insights into the effectiveness of our data transmission and recovery mechanisms. By leveraging Node MCU-generated ID, we ensure more reliable and efficient packet tracking and 
recovery, minimizing data loss and optimizing system performance.

\section{Conclusion}
Wireless Sensor Networks (WSNs) have revolutionized precision agriculture by enabling farmers to monitor and control agricultural processes with unprecedented accuracy and efficiency. In precision agriculture, accurate and synchronized data collection is crucial for effective analysis and decision-making. A pivotal challenge within this technological landscape is ensuring precise data synchronization among the numerous sensor nodes spread across vast agricultural fields. These issues can significantly impact the integrity and reliability of the data collected, essential for informed decision-making in agriculture. This mechanism enables the algorithm to adeptly handle node failures and maintain synchronization accuracy even in environments plagued by unpredictable communication conditions or intermittent connectivity. Reliable data storage is demanding now-a-days for storing data on local monitoring station as well as in online live server through internet. Sometime internet is not working properly due to congestion and there is frequent packet loss. Current solutions often synchronize records based on database timestamps, leading to record duplication and waste storage. Both databases synchronize each other after internet restoration. By providing synchronization among nodes and data, accuracy and storage will be saved in IoT based WSNs for precision agriculture applications. A prototype Node-MCU internal memory is used as a resource for achieving data synchronization.  Our proposed work generated record ID from Node MCU Electrically Erasable Programmable Read-Only Memory which helps in records synchronization if there is any packet loss at the local server or at the online server to maintain synchronization accuracy despite unreliable communication links. Experiment shows that for a particular duration Node MCU generated 2364 packets and packet loss at local server was 08 and at online server was 174 packets. Results shows that after synchronization 99.87\% packets were synchronized. Using previous technique of timestamp, the redundancy was 70\% which reduced to 0\% using our proposed technique.

This solution will surely bring many benefits to precision agriculture, especially in areas that have poor internet connectivity. Ensuring that data collection is accurate and reliable will help farmers make better decisions about crop management. This, in turn, has far-reaching implications for industries beyond agriculture, such as environmental monitoring and other IoT use cases enabled by real-time sensor data. The approach of synchronization through unique record IDs allows for a practical, scalable solution to meet the increasing demands for reliable data transmission within IoT networks.


 \bibliographystyle{alphaurl}
\bibliography{lmcs}



\end{document}